\documentclass[letterpaper]{article} 
\usepackage{aaai2026}  
\usepackage{times}  
\usepackage{helvet}  
\usepackage{courier}  
\usepackage[hyphens]{url}  
\usepackage{graphicx} 
\usepackage{natbib}  
\usepackage{caption} 
\usepackage{amsmath}
\usepackage{booktabs}
\usepackage{array} 
\usepackage{seqsplit}
\usepackage{algorithm}
\usepackage{algorithmic}

\usepackage{newfloat}
\usepackage{listings}
\DeclareCaptionStyle{ruled}{labelfont=normalfont,labelsep=colon,strut=off} 
\floatstyle{ruled}
\newfloat{listing}{tb}{lst}{}
\floatname{listing}{Listing}

\title{JW-Flare: Accurate Solar Flare Forecasting Method Based on Multimodal Large Language Models}
\author {
    Mingfu Shao\textsuperscript{\rm 1,2}\equalcontrib,
    Hui Wang\textsuperscript{\rm 1,2}\equalcontrib,
    Yuyang Li\textsuperscript{\rm 2,1}\equalcontrib,
    Jiaben Lin\textsuperscript{\rm 1,2}\equalcontrib\thanks{Corresponding author: jiabenlin@bao.ac.cn},
    Jifeng Liu \textsuperscript{\rm 1,2},
    Baolin Tan \textsuperscript{\rm 1,2},
    Juan Guo \textsuperscript{\rm 1,2},
    Yin Zhang \textsuperscript{\rm 1,2},
    Jing Huang \textsuperscript{\rm 1,2},
    Jiangtao Su \textsuperscript{\rm 1,2},
    Yingzi Sun \textsuperscript{\rm 1,2},
    Haiqing Xu \textsuperscript{\rm 1,2},
    Jie Chen \textsuperscript{\rm 1,2},
    Suo Liu \textsuperscript{\rm 1,2},
    Yuanyong Deng \textsuperscript{\rm 1,2},
    Liyue Tong \textsuperscript{\rm 1},
    Yang Bai \textsuperscript{\rm 1},
    Cunshi Wang \textsuperscript{\rm 2,1},
    Kaifan Ji \textsuperscript{\rm 3},
    Yuqing Zhou \textsuperscript{\rm 3,2}
}
\affiliations {
    \textsuperscript{\rm 1}State Key Laboratory of Solar Activity and Space Weather, NAOC, Beijing 100101, P. R. China;\\
    \textsuperscript{\rm 2}University of Chinese Academy of Sciences, Beijing 101408, P. R. China;\\
    \textsuperscript{\rm 3}Yunnan Observatories, Chinese Academy of Sciences, Kunming, 650216, China\\
    shaomf@bao.ac.cn
}

\nocopyright
\begin{document}

\maketitle

\begin{abstract}
Solar flares, the most powerful explosive phenomena in the solar system, may pose significant hazards to spaceborne satellites and ground-based infrastructure. Despite decades of intensive research, reliable flare prediction remains a challenging task. Large Language Models, as a milestone in artificial intelligence, exhibit exceptional general knowledge and next-token prediction capabilities. Here we introduce JW-Flare, the first Multimodal Large Language Models (MLLMs) explicitly trained for solar flare forecasting through fine-tuning on textual physic parameters of solar active regions and magnetic field images. This method demonstrates state-of-the-art (SOTA) performance for large flares prediction on the test dataset. It effectively identifies all 79 X-class flares from 18,949 test samples, yielding a True Skill Statistic (TSS) of 0.95 and a True Positive Rate (TPR) of 1.00, outperforming traditional predictive models. We further investigate the capability origins of JW-Flare through explainability experiments, revealing that solar physics knowledge acquired during pre-training contributes to flare forecasting performance. Additionally, we evaluate models of different parameter scales, confirming the Scaling\_Law of Large Language Models in domain-specific applications, such as solar physics. This study marks a substantial advance in both the scale and accuracy of solar flare forecasting and opens a promising avenue for AI-driven methodologies in broader scientific domains.
\end{abstract}


\section{Introduction}
Solar flares are the most intense explosions in the solar system and the primary drivers of catastrophic space weather events. These events can damage power grids and destroy satellites. In 2024, the Sun entered its solar maximum phase, marked by a significant increase in solar flare occurrences. Coinciding with this period, SpaceX's Starlink satellites experienced the highest number of reentry and damage events. Scientists from NASA analyzed over 500 Starlink satellites that prematurely deorbited between 2020 and 2024, finding that intense solar storms caused by powerful flare eruptions significantly increase electromagnetic radiation, which accelerates the orbital decay of low Earth orbit satellites such as Starlink, posing serious challenges to the sustainable utilization of near-Earth orbital space \cite{oliveira2025tracking}. Therefore, accurate forecasting of solar flares is crucial for mitigating societal and economic impacts.

\begin{figure}[t]
\centering
\includegraphics[width=0.9\columnwidth]{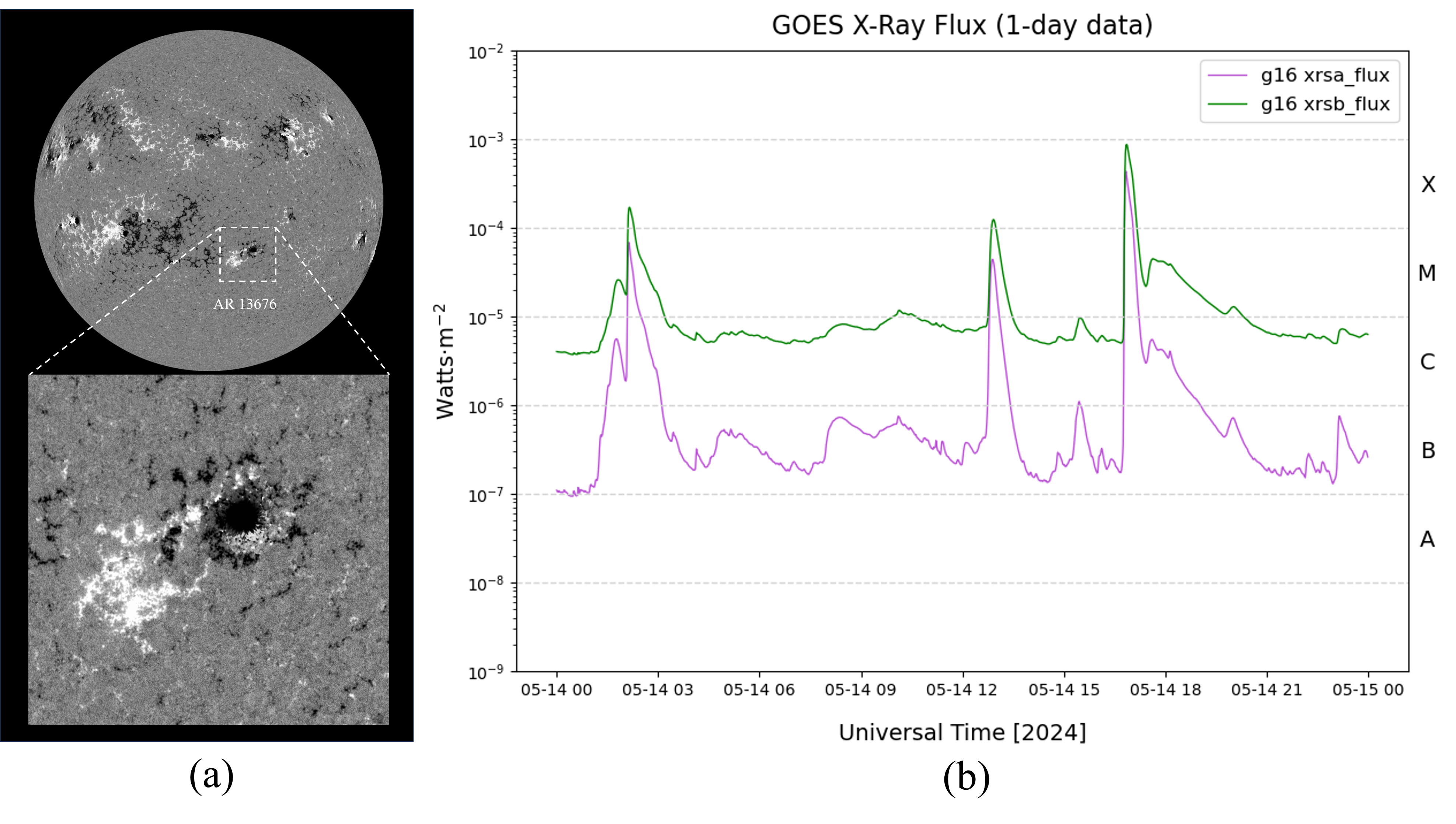} 
\caption{An X8.7-class solar flare erupted from AR13676 at 16:44~UT on May 14, 2024. (a) shows the line-of-sight magnetogram of the active region, while (b) presents the solar X-ray flux in the 0.5--4.0~\AA{} (\texttt{xrsa\_flux}) and 1.0--8.0~\AA{} (\texttt{xrsb\_flux}) bands, as measured by the X-Ray Sensor (XRS) onboard Geostationary Operational Environmental Satellite (GOES). Solar flares are categorized  into five intensity levels in decreasing order of intensity: X, M, C, B, and A, based on the xrsb\_flux.}

\label{fig1}
\end{figure}

Prior studies in solar physics have established a strong correlation between magnetic activity and flare eruptions, making Line-of-Sight (LoS) magnetograms a crucial input for flare forecasting models, as illustrated in Figure \ref{fig1}(a). In recent years, researchers have commonly relied on self-constructed datasets to perform solar flare forecasting within 24 hours. With the increasing availability of high-resolution observational data from instruments such as the Solar Dynamics Observatory/Helioseismic and Magnetic Imager (SDO/HMI), Boucheron et al. established a standardized dataset specifically designed for deep learning-based flare prediction. This dataset includes LoS magnetograms, derived magnetic parameters, and annotated flare labels indicating the peak class of flares occurring within the subsequent 24-hour window, covering the period from May 1, 2010 to December 31, 2018 \cite{boucheron2023solar}. Solar flare forecasting has evolved from statistical learning and traditional machine learning to modern deep learning approaches, predominantly framing the task as a data-driven classification problem to predict flare intensity classes. Nevertheless, although the growing adoption of data-driven artificial intelligence (AI) models in solar flare forecasting, accurately predicting higher-class flares remains a significant and unresolved challenge.

The recent proliferation of open-source large language models (LLMs), including Llama \cite{touvron2023llama}, DeepSeek \cite{guo2025deepseek}, and Qwen \cite{bai2023qwen}, has created unprecedented opportunities for scientific discovery. Advances in AI are rapidly reshaping the landscape of predictive science, enabling transformative progress in domains ranging from protein structure prediction \cite{jumper2021highly}, to climate modeling \cite{bi2023accurate}, and extreme weather forecasting \cite{camps2025artificial}. In particular, the emergence of multimodal large language models (MLLMs), capable of jointly processing visual data, textual information, and  numerical parameters, offering novel possibilities for cross-modal feature fusion and end-to-end predictive modeling. This paradigm shift holds significant promise for scientific fields characterized by rich multimodal data—such as solar physics—where integrating magnetograms, physical parameters, and flare annotations is essential for accurate flare forecasting.

Building on these advancements, we introduce JinWu-Flare (JW-Flare)—a novel AI-powered method for solar flare prediction that pioneers the use of MLLMs in this domain. The model is named after JinWu, the mythical solar bird in Chinese mythology, and its logo is presented in Figure 2. The experimental results demonstrate that JW-Flare achieves perfect Recall (100\% TPR) in identifying all 79 X-class flares within the 18,949-sample test dataset, while maintaining state-of-the-art (SOTA) performance in True Negative Rate (TNR) and TSS metrics. This work thus establishes a new benchmark in flare prediction through its innovative integration of MLLM technologies, offering both enhanced predictive accuracy and reduced implementation complexity compared to conventional deep learning frameworks.

In summary, Our main contributions are summarized as: 
\begin{itemize}

 \item The First MLLM for Solar Flare Prediction: To our knowledge, this is the first systematic application of MLLM to solar flare forecasting. This performance significantly outperforms conventional statistical and deep learning methods, underscoring the transformative potential of LLMs in scientific discovery.
 
 \item Explainability experiment and scaling\_law: We observe that domain-specific knowledge acquired during the pretraining phase of MLLMs contributes positively to the decision-making process in solar flare prediction. Furthermore, Scaling\_Law experiments confirm that the scaling\_laws of LLMs are equally valid and beneficial in scientific applications.

 \item Generalizable AI Paradigm: We introduce an ontology-reconfigurable framework that decouples domain-specific expertise from algorithmic complexity. By embedding complex computations into pre-trained MLLMs, our methodology enables researchers to focus on curating domain-specific prompts and datasets—an approach that can accelerate discovery in other interdisciplinary fields requiring complex reasoning.

\end{itemize}

This paper is organized as follows: Section 2 describes the related work of solar flare prediction. Section 3 describes the method of JW-Flare. Section 4 presents the results of experiment, assesses the generalization performance, explores the scaling law and explainability of the model. Additionally, it includes an evaluation of the effectiveness of the experimental design through ablation studies. Section 5 encapsulates the experimental findings, examines the limitations inherent in the current study and proposes prospective avenues for future enhancement.

\section{Related Work}

The first recorded solar flare occurred on September 1, 1859, and was independently observed by Richard C. Carrington and Richard Hodgson. In 1939, Giovanelli et al. \cite{giovanelli1939relations} initiated the first systematic investigation into the  relationship between sunspots and solar flares. Over the past century, research on solar flare forecasting has been primarily categorized into physics-based methods and data-driven methods. Physics-based methods focus on uncovering the underlying mechanisms of flare eruptions, while data-driven methods aim to establish forecasting models through the analysis of extensive observational data. 

\subsection{physics-based methods}

Since the discovery of solar flares, solar physicists have devoted considerable effort to elucidating the physical origins of eruptions and constructing effective predictive frameworks \cite{huang2024short}. For example, Self-Organized Criticality (SOC) models, derived from first principles, have been widely explored to simulate the intricate processes of solar evolution and flare formation by employing numerical simulations to predict the spatial distribution and temporal evolution of key physical parameters \cite{karakatsanis2008soc}. However, the inherent complexity of solar flare mechanisms has prevented a definitive consensus on their physical origins and continues to pose significant challenges to developing precise physics-based predictive models. \cite{korsos2018applying, lin2009studies, ning2009investigation, ning2012power, wang2012solar}.

\subsection{data-driven methods}

\textbf{Statistical methods:}
As solar observational data continue to accumulate, data-driven approaches for solar flare prediction have evolved significantly in methodology. Initially, the Space Environment Laboratory of National Oceanic and Atmospheric Administration, in collaboration with the University of Colorado, developed the first expert system to predict solar flares by utilizing the relationship between McIntosh Sunspot Classifications and flare activity. Later on, Bloomfield et al. \cite{bloomfield2012toward}proposed a probabilistic prediction model based on a Poisson distribution, which estimated the likelihood of different flare classes by calculating the average flare eruption rate for within each individual McIntosh sunspot class. At that time, solar flare prediction primarily relies on domain experts manually identifying key predictive features (e.g., magnetic field parameters and sunspot classifications) and employing statistical learning methods to model empirical relationships between these features and flare events.

\textbf{Machine learning methods:}
The emergence of machine learning frameworks has effectively mitigated the limitations of traditional statistical learning by enabling semi-automated feature extraction, in which domain experts define the feature space and directional constraints, while algorithms iteratively optimize these features. Yuan et al. (2016) \cite{yuan2020solar} applied Principal Component Analysis (PCA) to extract feature from sunspot active region parameters(e.g., shape, area, and other morphological features) and 10.7 cm radio flux, which were subsequently used as inputs for an Support Vector Machines(SVM) prediction model. Ensemble approaches further enhanced predictive robustness, as shown in Abduallah et al. (2021) \cite{abduallah2021deepsun}, which integrated multiple learning algorithms through a voting mechanism to predict flares of varying magnitudes. The reliance on manual intervention for feature selection during this period posed significant challenges for automated forecasting, necessitating extensive data preprocessing to extract meaningful predictors from raw observations.

\textbf{Deep learning methods:}
In contrast, deep learning leverages multi-layer nonlinear architectures to automatically learn hierarchical features, significantly reducing the reliance on manual feature selection and enabling fully automated feature extraction. Huang et al. (2018) \cite{huang2018deep} employed Convolutional Neural Network (CNN) to automatically identify regions with opposing magnetic polarities, eliminating the need for manual feature engineering. Nishizuka et al. \cite{nishizuka2018deep, nishizuka2020reliable, nishizuka2021operational} applied deep neural networks to analyse 79-dimensional physical features extracted from active region magnetograms, aiming to predict 24-hour solar flare probabilities. Their results demonstrated that such models are capable of capturing complex spatiotemporal patterns inherent in high-dimensional magnetogram data. Zheng et al. (2023) \cite{zheng2023multiclass} proposed the HBiLSTM-Attention model, which integrates an attention mechanism to enhance multi-class flare prediction by leveraging the spatial distribution and temporal evolution of solar active region (AR) magnetograms. Similarly, Abduallah et al. (2023) \cite{abduallah2023operational} developed SolarFlareNet, a hybrid model integrating CNN, Long Short-Term Memory (LSTM), and Transformer architectures, using magnetic field parameters extracted from the Spaceweather Helioseismic and Magnetic Imager Active Region Patch (SHARP) data to predict flares with intensities of $\geq$M5.0-class, $\geq$M-class,and $\geq$C-class, respectively. While these approaches enable automated feature extraction, their dependence on single-modality data (e.g., images, textual descriptions, or univariate time series) inherently restricts their capacity to model cross-modal interactions and integrate contextual information.

\section{Method}
In recent years, MLLMs have demonstrated remarkable predictive performance by effectively integrating heterogeneous modalities such as images and text. While LLMs possess broad general-purpose language understanding acquired through unsupervised learning, their direct application to specialized scientific tasks (e.g., solar flare prediction) often yields suboptimal results. To address this limitation, we employ a supervised fine-tuning (SFT) approach \cite{touvron2023llama2openfoundation} to adapt general-purpose LLMs to the specific data characteristics of solar physics. The overall architectural workflow is illustrated in Figure 2. The framework first curates open-source data and injects prompt templates to construct a SFT dataset with balanced class distribution. It then employs LoRA technique to efficiently fine-tune the decoder parameters, utilizing a dual-input mechanism to separately encode magnetogram features from active regions and corresponding textual physical parameters, followed by cross-modal semantic alignment and fusion within deep layers. The model ultimately generates predictions under strict classification constraints (e.g., "A: Flare" or "B: None").

\begin{figure*}[!htb]
\centering
\includegraphics[width=0.9\textwidth]{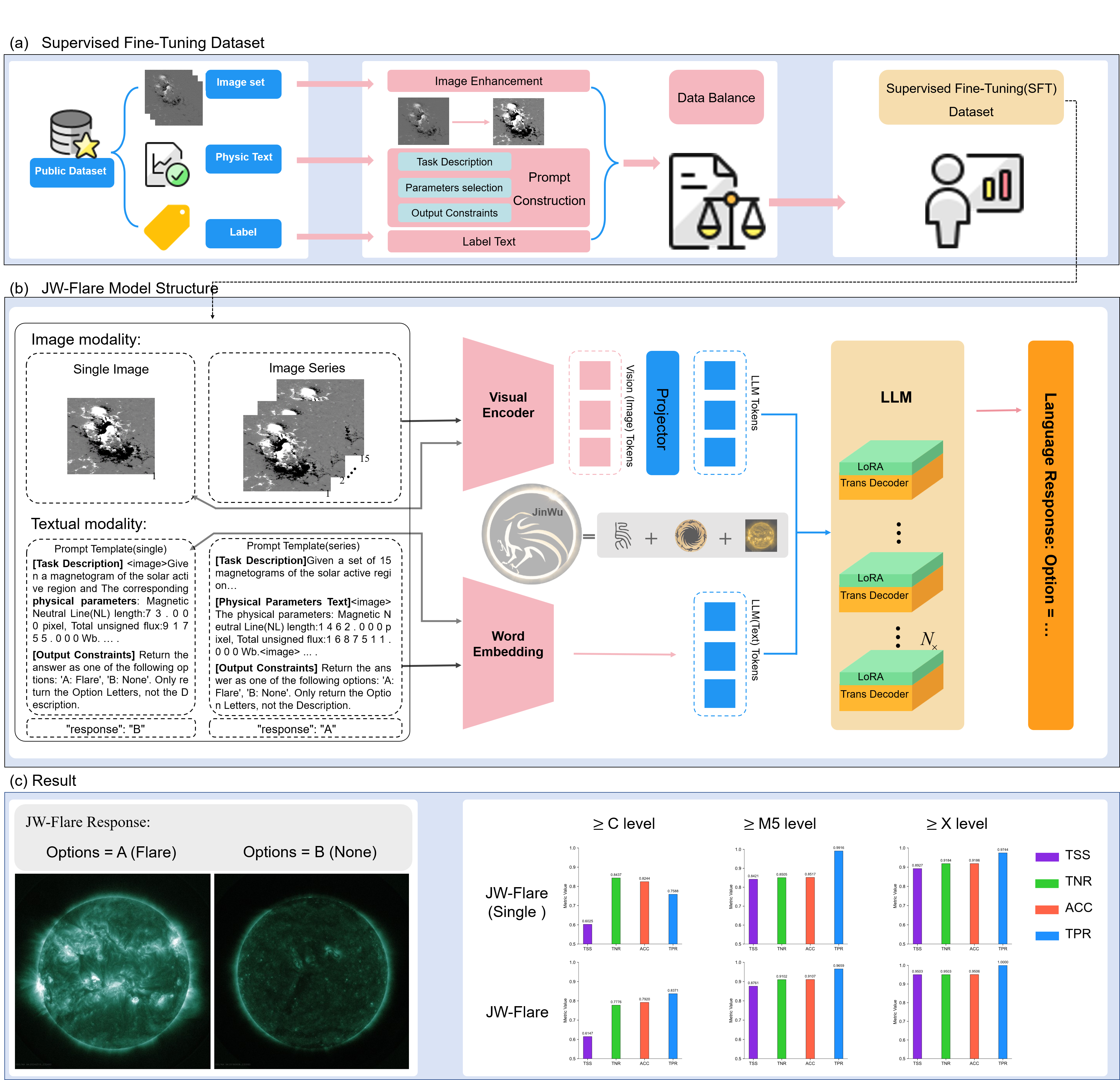} 
\caption{The flowchart of JW-Flare. (a) illustrates the flowchart of the SFT Dataset Construction. (b) illustrates the model structure of JW-Flare. The JinWu logo synthesizes three distinct cultural and scientific elements: The ancient Chinese seal script character: "\raisebox{-0.5ex}{\includegraphics[scale=0.01]{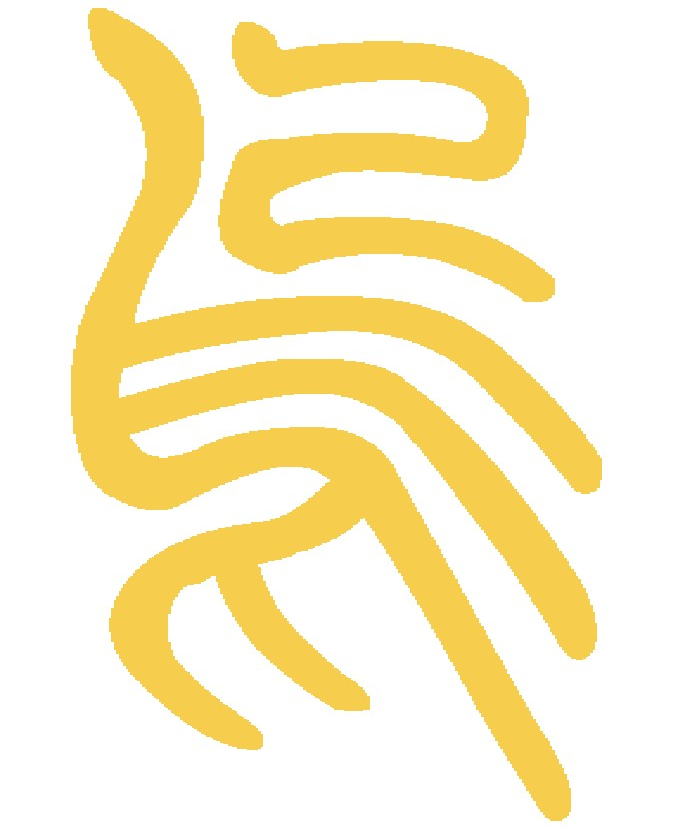}}"; the golden Sun Bird motif -- a culturally significant artifact unearthed at the Jinsha Site (circa 1200-650 BCE) representing solar worship; SDO's Atmospheric Imaging Assembly (AIA) 171 Å band imagery. (c) showcases the experimental results, with both solar images derived from the SDO/AIA 94 Å band. JW-Flare (single) represents the comparative experiment using a single image and its corresponding text as input.}
\label{fig2}
\end{figure*}

\subsection{Supervised Fine-Tuning dataset}
Compared to the prohibitively high computational costs required for training LLMs from scratch, SFT approaches enable rapid adaptation to the solar flare forecasting task by modifying only a small subset of model parameters, thereby achieving efficient development under resource-efficient setup. Due to inconsistencies in annotation formats and lack of contextual information in existing datasets within the domain, direct adaptation to the training requirements of MLLMs remains challenging. Therefore, constructing a task-specific SFT dataset for solar flare perdiction is essential to improve model accuracy and generalization. Figure 2(a) illustrates the process of constructing a customized SFT dataset for LLMs, based on the publicly available dataset released by Boucheron et al. \cite{boucheron2023solar}. 

\textbf{Image enhancement}: Since high-definition and high-contrast images can significantly improve the MLLM's accuracy in identifying solar active region features, we enhanced the sharpness and contrast of LoS magnetograms by applying a threshold optimization algorithm to the Flexible Image Transport System (FITS) file  provided by Boucheron et al. The pixel intensity values are preprocessed by first clipping them to the interval $[-200, 200]$, followed by normalization to the range $[0,255]$, which results in high-quality \(600 \times 600\) pixel PNG images suitable for model input. The specific enhancement strategy is outlined in Equation (1):
\begin{equation}
  I = \left[ \mathrm{MinMax}(0, 400, i + 200) \times \frac{255}{400} \right]
\end{equation}
Where \text{MinMax}(0, 400, i + 200) represents clipping the data intensity `i` to the range $[-200, 200]$, and the final normalized image is in the range $[0, 255]$.

\textbf{Textual construction}: 
To construct a domain-specific SFT dataset for JW-Flare, we enable efficient multimodal fusion of magnetograms and associated physical parameters by formulating prompt-based textual inputs, as illustrated in  Figures 2(a). This approach eliminates the need for complex data alignment or feature engineering, highlighting the flexibility and practicality of prompt-based multimodal modeling in domain adaptation. 
Based on the structured prompt framework proposed by Li et al. \cite{Li_2025}, we have developed a solar flare prediction textual prompt template with a triple-guidance mechanism: the task description explicitly defines the model's requirement for magnetic analysis of  active regions and make decisions based on statistical regularities. In the model, physical parameters establish the connection between image and text data, and can be processed simultaneously to achieve multimodal alignment. Specifically, we selected two magnetic parameters strongly correlated with flare activity: magnetic neutral line length, which quantifies the spatial extent of polarity inversion regions, and total unsigned magnetic flux, a robust indicator of active-region magnetic energy storage \cite{nishizuka2018deep, chen2021flare}; the output constraints the model to return binary classification results, effectively mitigating the common issue of ambiguous expressions in generative models. Through Textual prompt construction, JW-Flare gains the capability for solar flare prediction and achieves strong performance.

\begin{table}[t]
\centering
\begin{tabular}{lllll}
    \toprule
    Model & Level & Train & Test & Val \\
    \midrule
    & $\geq$ C & 603328 & 94757 & 95933 \\
    JW-Flare(single) & $\geq$ M5.0 & 153976 & 94757 & 95933 \\
    & $\geq$ X & 28726 & 94757 & 95933 \\
    & $\geq$ C & 121128 & 18582 & 18822 \\
    JW-Flare & $\geq$ M5.0 & 28726 & 18582 & 18822 \\
    & $\geq$ X & 30219 & 18949 & 19184 \\
    \bottomrule
\end{tabular}
\caption{Sample distributions of the SFT training, validation, and test sets for JW-Flare in predicting $\geq$C, $\geq$M5, and $\geq$X-class solar flares. JW-Flare (single) compares inputs of a single image and coupled text.}
\label{table1}
\end{table}

\textbf{Data balance}: To mitigate severe class imbalance (flare vs. non-flare samples) caused by the low daily occurrence rate of solar flares, we implemented a combination of oversampling minority class instances and undersampling majority cases \cite{wan2023flare, liu2023selective}, achieving a balanced dataset with a 1:1 class ratio. In Boucheron's dataset, 950,047 magnetograms were divided into training, validation, and test sets in a 15:2:2 ratio. The sample ratio of flare/non-flare in the training set is approximately 1:4. The proportions in the validation and test sets remain unchanged to preserve the model’s ability to forecast solar flares in real-world scenarios. Table 1 shows the sample distribution of the SFT dataset corresponding to each flare level. The specific oversampling and undersampling ratios for each flare intensity level in train sets are as follows:

\begin{equation}
\begin{aligned}
(\geq \mathrm{C}\times2):\left(\frac{\text{Non-flare}}{2}\right)
&= (\geq \mathrm{M5.0}\times14):\left(\frac{\text{Non-flare}}{10}\right)\\
\llap{=}\,(\geq \mathrm{X}\times30):\left(\frac{\text{Non-flare}}{10}\right)
&=\;\;\;\;\;\;\;\;\;\;\;\;\;\;\; 1:1
\end{aligned}
\end{equation}

A time-series input sample comprising a sequence of 15 images (captured at 12-minute intervals) and their associated textual parameters, captured over a 3-hour observation window, was utilized for JW-Flare to forecast  flare over the subsequent 24-hour period. We had also compared the performance of JW-Flare using single image and coupled text as input. Table 2 presents the complete SFT sample of JW-Flare.

\begin{table}[t]
\centering
\begin{tabular}{p{8cm}} 
\toprule
        \textbf{SFT sample of JW-Flare: } \\
        \midrule
        \textbf{query}: "Given a set of 15 magnetograms ..., please follow the steps below to conduct a step-by-step analysis...: Step 1: ...Step 2: ...Step 3: ...Step 4: .... \texttt{<image>}The physical parameters: Magnetic Neutral Line(NL) length:1~4~6~2~.~0~0~0 pixel, Total unsigned flux:1 6 8 7 5 1 1~.~0~0~0 Wb.\texttt{<image>}... ....Return the answer as one of the following options: 'A: Flare', 'B: None'. Only return the Option Letters, not the Description.", \\ 
        \textbf{response}: "A", \\ 
        \textbf{history}: [], \\ 
        \textbf{images}: ["/***.png" ...] \\ 
        \bottomrule
\end{tabular}
\caption{Dataset Example of JW-Flare. The model's inputs include 'query', which refers to the prompt template; 'history', which refers to the record of previous dialogue exchanges; and 'images', which represent the paths to the images included within the query and correspond one-to-one with the \texttt{<image>} tags. "response" denotes the corresponding labels output by the model, ‘A’ represents flare, and ‘B’ represents no flare.}
\label{table2}
\end{table}

\subsection{Model structure}

JW-Flare is built upon the supervised fine-tuning of Qwen2-VL-7B-Instruct, an open-source MLLM developed by TongYi, enabling effective multimodal learning for solar flare prediction, as shown in Figure 2(b) \cite{wang2024qwen2}. We first present the SFT dataset, which integrates textual and image data modalities. The textual modality comprises textual prompt templates (containing task descriptions, magnetic field parameters and output constraints) and corresponding flare labels, while the image modality consists of LoS magnetograms of active regions acquired from the SDO/HMI. The Visual Encoder employing a Vision Transformer (ViT) \cite{dosovitskiy2020image} architecture first partitions the image into patches and subsequently converts them into a variable number of visual tokens. A Projection layer aligns the vector dimensions of visual tokens with text tokens processed through Word Embedding. Next, both of them are concatenated within the embedding space and fed into the LLM. 

To enable efficient fine-tuning, we applied the LoRA technique \cite{hu2021lora}. Specifically, we froze the pre-trained weights of both the ViT encoder and the projection layer, while keeping the LLM fully active. Within the LLM, all linear layers of the Transformer modules were trained using low-rank adaptation matrices. This approach significantly reduces the number of trainable parameters and computational costs compared to full-parameter fine-tuning, while preserving predictive performance.

The model generates its final output in the form of predefined options, as specified in the prompt text. JW-Flare uses a 3-hour time-series sample (comprising 15 images) to forecast flares over the next 24 hours. In contrast, the JW-Flare(single) variant predicts flares occurring 24 hours ahead using only a single magnetogram image paired with associated textual parameters.

\section{Results}
The performance of JW-Flare is evaluated through comprehensive experiments, encompassing comparisons with baseline methods and rigorous assessments of the model’s generalization ability. Furthermore, we investigate the model’s explainability and examine the applicability of scaling laws within the scientific domain. Finally, ablation studies are conducted to evaluate the impact of various architectural and design choices.

\subsection{Experimental settings}
The experiments were conducted on a high-performance computing platform running a Linux-based operating system equipped with four NVIDIA RTX 6000 Ada GPUs. The experimental framework is built on PyTorch and ModelScope Swift — a lightweight toolkit designed for efficient fine-tuning of large-scale models \cite{zhao2024swift}. Leveraging the LoRA technique, all models were fine-tuned under low-resource settings using a single NVIDIA RTX 6000 Ada GPU, achieving efficient training of 7-billion-parameter MLLMs. Training $\geq$C-class flare prediction model took approximately 52 hours per GPU, compared to around 15 hours for $\geq$M5.0/X-class models. 

We evaluate JW-Flare using standard classification metrics. Specifically, we report the True Positive Rate (TPR or Recall), Precision, and the True Negative Rate (TNR). Accuracy (ACC) is informative for balanced data; however, for imbalanced scenarios such as solar flare forecasting, metrics including the True Skill Score (TSS), F1 Score, and Heidke Skill Score (HSS) provide a more robust assessment. Among these metrics, the True Skill Score (TSS) is considered the most essential for performance evaluation.

Figures 3 depicts the loss curve and the learning rate curve for JW-Flare. The downward trend in both curves signifies effective convergence, with the model employing a cosine annealing learning rate schedule. In this system, the learning rate increases gradually in the early stages of training and then decays over time. This strategy facilitates a more thorough exploration of the parameter space during the initial phase and allows for refined optimization in the later stages, thereby enhancing the model's convergence performance.

\begin{figure}[!htbp]
  \centering
  \includegraphics[width=0.9\columnwidth]{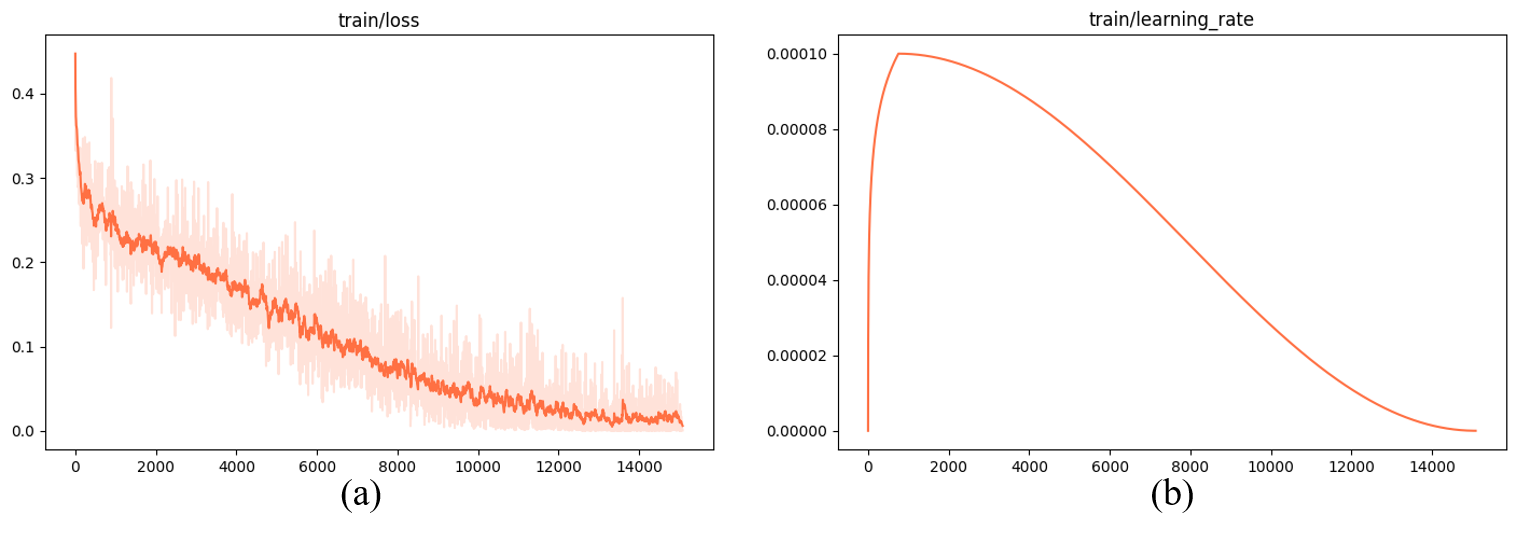}
  \caption{The loss curve (a) and learning rate curve (b) for JW-Flare, specifically for predictions of C-class and largher solar flares.}
  \label{fig3}
\end{figure}

\subsection{JW-Flare's experiment results}
The standardized output format of JW-Flare and its performance evaluation metrics, derived from the SFT test dataset, are shown in Figure 2(c).  For any given AR, JW-Flare has the capability to predict whether a flare of a specific magnitude will occur within the next 24 hours. Specifically, it separately forecasts flares of $\geq$X-class ($10^{-4} \text{Wm}^{-2}$), $\geq$M5.0-class ($5\times10^{-5} \text{Wm}^{-2}$), and $\geq$C-class ($10^{-6} \text{Wm}^{-2}$), which are critical for space weather monitoring. Table 3 provides a detailed summary of the model's performance metrics, with the TSS highlighted in bold. JW-Flare demonstrated peak performance in predicting X-class flares, correctly identifying all 79 X-class "Flare" samples without any missed detections, achieving an exceptional TSS value of 0.9503. Moreover, the model accurately predicted 17,933 flare-free events, with only 937 false positives. Among these, 290 samples were found to correspond to M-class flares, 553 to C-class flares, and just 94 were truly flare-free events, underscoring the model's high prediction accuracy for X-class flares. While some false positives were detected, approximately 90\% of these misclassified samples were associated with flare of lower intensities than X-class. In addition, JW-Flare achieved TSS values of 0.8761 for predicting M5.0-class and larger flares, representing the highest values reported to date. The experimental results demonstrate that JW-Flare exhibits significantly superior predictive capabilities for higher-class flares ($\geq$M5-class and $\geq$X-class) compared to $\geq$C-class flares, with key evaluation metrics (e.g., TSS, TNR, ACC, TPR) showing notable improvements. Notably, JW-Flare achieves superior performance with sequential image inputs compared to JW-Flare(single), as MLLM leverages temporal modeling capability to analyze the evolutionary trends of active regions in the image sequence, enabling more accurate flare eruption discrimination, particularly in the TSS metric. 

\begin{table*}[!htb]
  \centering
  \renewcommand{\arraystretch}{2}
  \setlength\tabcolsep{1.5pt}
  \label{tab:jw-flare-performance}
  \small
  \begin{tabular}{ccccccccccccc}
      \toprule
      Model & Level & TP & TN & FP & FN & TPR & TNR & HSS & ACC & Precision & F1 & \textbf{TSS} \\
      \midrule
      JW-Flare (single) & $\geq$ C & 16345 & 61772 & 11443 & 5197 & 0.7588 & 0.8437 & 0.5465 & 0.8244 & 0.5882 & 0.6627 & \textbf{0.6025} \\
      JW-Flare & $\geq$ C & 3767 & 10950 & 3132 & 733 & 0.8371 & 0.7776 & 0.5203 & 0.7920 & 0.5460 & 0.6609 & \textbf{0.6147} \\
      JW-Flare (single) & $\geq$ M5 & 826 & 79879 & 14045 & 7 & 0.9916 & 0.8505 & 0.0900 & 0.8517 & 0.0555 & 0.1052 & \textbf{0.8421} \\
      JW-Flare & $\geq$ M5 & 170 & 16753 & 1653 & 6 & 0.9659 & 0.9102 & 0.1555 & 0.9107 & 0.0933 & 0.1701 & \textbf{0.8761} \\
      JW-Flare (single) & $\geq$ X & 342 & 86700 & 7706 & 9 & 0.9744 & 0.9184 & 0.0749 & 0.9186 & 0.0425 & 0.0814 & \textbf{0.8927} \\
      JW-Flare & $\geq$ X & 79 & 17933 & 937 & 0 & 1.0000 & 0.9503 & 0.1376 & 0.9506 & 0.0778 & 0.1443 & \textbf{0.9503} \\
      \bottomrule
  \end{tabular}
  \caption{Performance of JW-Flare on the SFT test dataset: The bolded metrics represent key performance indicators, where larger values indicate stronger model performance.The significance of specific evaluation metrics can be referred to in the appendix A.2.}
\end{table*}

\subsection{Comparative analysis based on the public dataset}
In order to provide a comprehensive evaluation of JW-Flare's performance, we conduct a comparative analysis using the Public Dataset released by Boucheron. Table 4 presents the comparison results in predicting C-class and larger flares within the next 24 hours. All models use Boucheron's dataset (FITS files) for training and testing. Baseline performance metrics for the SVM \cite{cortes1995support} and Visual Geometry Group (VGG) \cite{simonyan2014very} models are provided by Boucheron et al. The SVM model utilizes physical features for binary flare classification, while the VGG model directly reads the FITS files and linearly scales them to [0, 255] for binary classification. To further assess the effectiveness of JW-Flare, we also introduce the Swin Transformer \cite{liu2021swin}, which was trained and tested on the same dataset as the VGG model. Moreover, JW-Flare incorporates an enhanced preprocessing strategy to improve image contrast and clarity, achieving the highest TSS value of 0.6147. The JW-Flare (single) model, which uses a single image as input, also attains a high TSS of 0.6025. The experimental results demonstrate that JW-Flare excels in flare forecasting, underscoring the effectiveness and superiority of the MLLM in solar flare prediction.

\begin{table*}[!htb]
  \centering
  \renewcommand{\arraystretch}{2}
  \setlength\tabcolsep{1.5pt}
  \label{tab:jw-flare}
  \small
  \begin{tabular}{ccccccccccccc}
      \toprule
      Model & Level & TP & TN & FP & FN & TPR & TNR & HSS & ACC & Precision & F1 & \textbf{TSS} \\
      \midrule
      SVM   & $\geq C$ & 16123 & 57042 & 16173 & 5419  & 0.7484 & 0.7791 & 0.4485 & 0.7721 & 0.4992 & 0.5989 & 0.5275 \\
      VGG   & $\geq C$ & 14977 & 59611 & 13604 & 6565  & 0.6952 & 0.8142 & 0.4567 & 0.7872 & 0.5240 & 0.5976 & 0.5094 \\
      Swin\_transformer & $\geq C$ & 16186 & 58927 & 14288 & 5356  & 0.7514 & 0.8048 & 0.4852 & 0.7927 & 0.5311 & 0.6223 & 0.5562 \\
      JW-Flare (single) & $\geq$ C & 16345 & 61772 & 11443 & 5197 & 0.7588 & \textbf{0.8437} & \textbf{0.5465} & \textbf{0.8244} & \textbf{0.5882} & \textbf{0.6627} & 0.6025 \\
      JW-Flare & $\geq$ C & 3767 & 10950 & 3132 & 733 & \textbf{0.8371} & 0.7776 & 0.5203 & 0.7920 & 0.5460 & 0.6609 & \textbf{0.6147} \\
      \bottomrule
  \end{tabular}
  \caption{Comparison based on the Boucheron's Dataset. The highest performance metrics are highlighted in bold.}
\end{table*}

\subsection{Comparative analysis with existing solar flare prediction models}

We evaluate multiple solar flare prediction models within the same 24-hour forecasting window, where JW-Flare demonstrates the best overall performance. Notably, JW-Flare (single) is designed to accept a single image and its associated textual information, allowing for direct comparison with the models listed in Table 5 under identical input conditions. 

\begin{table*}[!htb]
  \centering
  \renewcommand{\arraystretch}{2}
  \setlength\tabcolsep{1.5pt}
  \label{tab:jw-flare-per}
  \tiny
  \begin{tabular}{cccccccccccc}
      \toprule
      Method & Model & Type & Input & Level  & Table & TPR & TNR & ACC & HSS & \textbf{TSS}\\
      \midrule
      Huang et al. (2018)\cite{huang2018deep} & CNN & AR.C 2 & Single & C/M/X  & T4 & 0.73/0.85/0.87 & 0.76/0.81/0.85 & 0.76*/0.81*/0.85* & 0.34/0.14/0.03 & 0.49/0.66/0.71\\
      Nishizuka et al. (2018)\cite{nishizuka2018deep} & DNN & AR.F 1* & Single & C+/M+  & T3 & 0.81/0.95 & 0.82*/0.85* & 0.82/0.86 & 0.53/0.26 & 0.63/0.80\\
      Park et al. (2018)\cite{park2018application} & CNN & FD 2 & Single & C+/*  & T4 & 0.80* & 0.89* & 0.84 & --- & 0.69\\
      Li et al. (2020)\cite{li2020predicting} & CNN & AR.C 2 & Single & C+/M+  & T2 & 0.89/0.82 & 0.79*/0.93* & 0.86/0.89 & 0.67/0.76 & 0.68/0.75\\
      Nishizuka et al. (2021)\cite{nishizuka2021operational} & DNN,t=0.5 & AR.F 1* & Single & C+/M+  & T2,T3 & 0.71*/0.25 & 0.99*/0.99* & 0.99/0.99 & 0.64/0.06 & 0.70/0.24\\
      Zheng et al. (2021)\cite{zheng2021hybrid} & CNN & AR.C 2 & Single & C+/M+  & T3 & 0.90/0.82 & 0.81*/0.92* & 0.87/0.89 & 0.69/0.75 & 0.70/0.74\\
      Deng et al. (2021)\cite{deng2021fine} & GAN+CNN & AR.C 2 & Single & C/M/X  & T8 & 0.73/0.75/0.83 & 0.91*/0.91*/0.93* & 0.87/0.86/0.91 & 0.64/0.66/0.77 & 0.65/0.65/0.76\\
      Sun et al. (2022a)\cite{sun2022solar} & 3D CNN & AR.C 3 & Series & C+/M+  & T5 & 0.86/0.93 & 0.89*/0.90* & 0.88/0.90 & 0.76/0.67 & 0.76/0.83\\
      Abduallah et al. (2023)\cite{abduallah2021deepsun} & Transformer & AR.C 1 & Series & C+/M+/M5+  & T1 & 0.89/0.84/0.85 & 0.94*/0.98/0.97* & 0.92/0.93/0.96 & --- & 0.84/0.84/0.82\\
      JW-Flare(single) & MLLM & AR.C 2 & Single & C+/M5+/X  & T2 & 0.76/0.99/0.97 & 0.84/0.85/0.92 & 0.82/0.85/0.92 & 0.55/0.09/0.08 & 0.60/0.84/0.89\\
      JW-Flare & MLLM & AR.C 3 & Series & C+/M5+/X  & T2 & 0.84/0.97/1.00 & 0.78/0.91/0.95 & 0.79/0.91/0.95 & 0.52/0.16/0.14 & 0.61/0.88/0.95\\
      \bottomrule
  \end{tabular}
  \caption{Comparison of different Solar Flare Prediction Methods. Model: Refers to the foundation model for various methods. Type: AR.C uses central active region data; AR.F includes data from all active regions; FD represents full-disk data. Numbers (1, 2, 3) indicate data dimensionality: 1D textual parameter, 2D image, or 3D time-sequence images. Input: Indicates the type of data input, with "single" representing a single image or parameter, and "series" representing a time sequence of images or parameters. Level: Refers to the class of flares predicted by the model. Table: Corresponding table from the original paper. All the evaluation metrics were extracted from the original paper. ‘---'indicates unavailable data, ’*’ indicates data that were not explicitly provided in the original articles but were inferred through our analysis.}
\end{table*}

Taking MLLM as the foundation model, JW-Flare offers significant advantages in terms of data compatibility, supporting a diverse range of inputs, including single image, time-series images, and active region magnetic parameters. This flexibility greatly enhances the data diversity for flare prediction, allowing the model to better capture various aspects of solar activity. As shown in Table 4, Through rigorous comparisons with a wide range of methods, JW-Flare demonstrates remarkable performance, achieving a TSS of 0.62 for $\geq$C-Class flares, rising to 0.88 for $\geq$M5.0-Class flares, and reaching 0.95 for $\geq$X-Class flares. Especially in forecasting $\geq$M5.0 and $\geq$X-Class flares, JW-Flare consistently surpasses previous models by approximately 10\% percentage points. Besides, accurately identifying true positive samples is crucial and JW-Flare excels achieving a 100\% TPR in predicting large flares. Overall, JW-Flare fully exploits the MLLM's capability to integrate multimodal information, achieving SOTA performance in predicting high-magnitude flares that are critical for space weather operations.

\subsection{Generalization capabilities test}
We conducted generalization tests using data observed by the 35CM magnetic field telescope at HuaiRou Solar Observing Station (HSOS) from March 2022 to December 2023, Advanced Space-based Solar Observatory/Full-disc vector MagnetoGraph (ASO-S/FMG) data from May 2024, and SDO/HMI data spanning from January to June 2024. JW-Flare(single) was evaluated on positive samples with flares from HSOS and FMG, achieving TPR values of 0.83 and 0.99, respectively; JW-Flare was evaluated on HMI samples, yielding a TPR of 0.71 and a TSS of 0.64, which indicate a significant decline in performance relative to the SFT test dataset. This decline can be attributed to the substantial temporal gap between the training and test periods, as JW-Flare’s training data spans from 2010 to 2018, while the HMI test samples are more recent. Additionally, differences in solar activity levels and flare occurrence probabilities across these periods further contribute to the reduced performance. While showing performance degradation in generalization tests, JW-Flare achieves reasonable accuracy (ACC=0.92) with acceptable true positive rates (TPR=0.71) on independent 2024 data, indicating its potential for real-world application. Future work will enhance model reliability through expanded temporal coverage in training data.

\begin{table*}[!htb]
  \centering
  \renewcommand{\arraystretch}{2}
  \setlength\tabcolsep{1.5pt}
  \label{tab:jw}
  \small
  \begin{tabular}{cccccccccccccc}
      \toprule
      Method & Data & Input & TP & TN & FP & FN & TPR & TNR & HSS & ACC & Precision & F1 & TSS \\
      \midrule
      JW-Flare(single) & HSOS & Single & 400 & 0 & 0 & 83 & --- & --- & --- & 0.828 & --- & --- & --- \\
      JW-Flare(single) & FMG & Single & 289 & 0 & 0 & 2 & --- & --- & --- & 0.993 & --- & --- & --- \\
      JW-Flare & HMI & Series & 236 & 21852 & 1732 & 95 & 0.713 & 0.927 & 0.186 & 0.924 & 0.120 & 0.205 & 0.640 \\
      \bottomrule
  \end{tabular}
  \caption{Generalization Capabilities Test.‘---’ indicates unavailable data}
\end{table*}

\subsection{Explainability experiment}

The explainability of AI methods, which is of great importance to scientific discovery, remains a widely discussed issue. Here, we explore why MLLM can accurately predict solar flares. Our approach is based on the  training method proposed by Muennighoff et al. \cite{muennighoff2025s1simpletesttimescaling} and knowledge distillation framework of the DeepSeek-R1 model \cite{guo2025deepseek}, where we extract textual modality data (physical parameters) from Boucheron's dataset—including Gradient mean, Total unsigned flux, Magnetic Neutral Line (NL) length, and Number of fragments along the magnetic NL (NL no. fragments)—to construct a Chain-of-Thought (CoT) dataset for exploring the reasoning mechanisms of LLMs. Furthermore, we employ neuron activation tracking technology \footnote{Monitor project: \url{https://monitor.transluce.org/dashboard/chat}} to conduct explainability analysis of the model's decision-making process \cite{choi2024automatic} (Physical interpretation of the parameters and the whole CoT reasoning process of the LLM can be found in Appendix).

\begin{figure*}[!htb]
  \centering
  \includegraphics[width=0.9\linewidth]{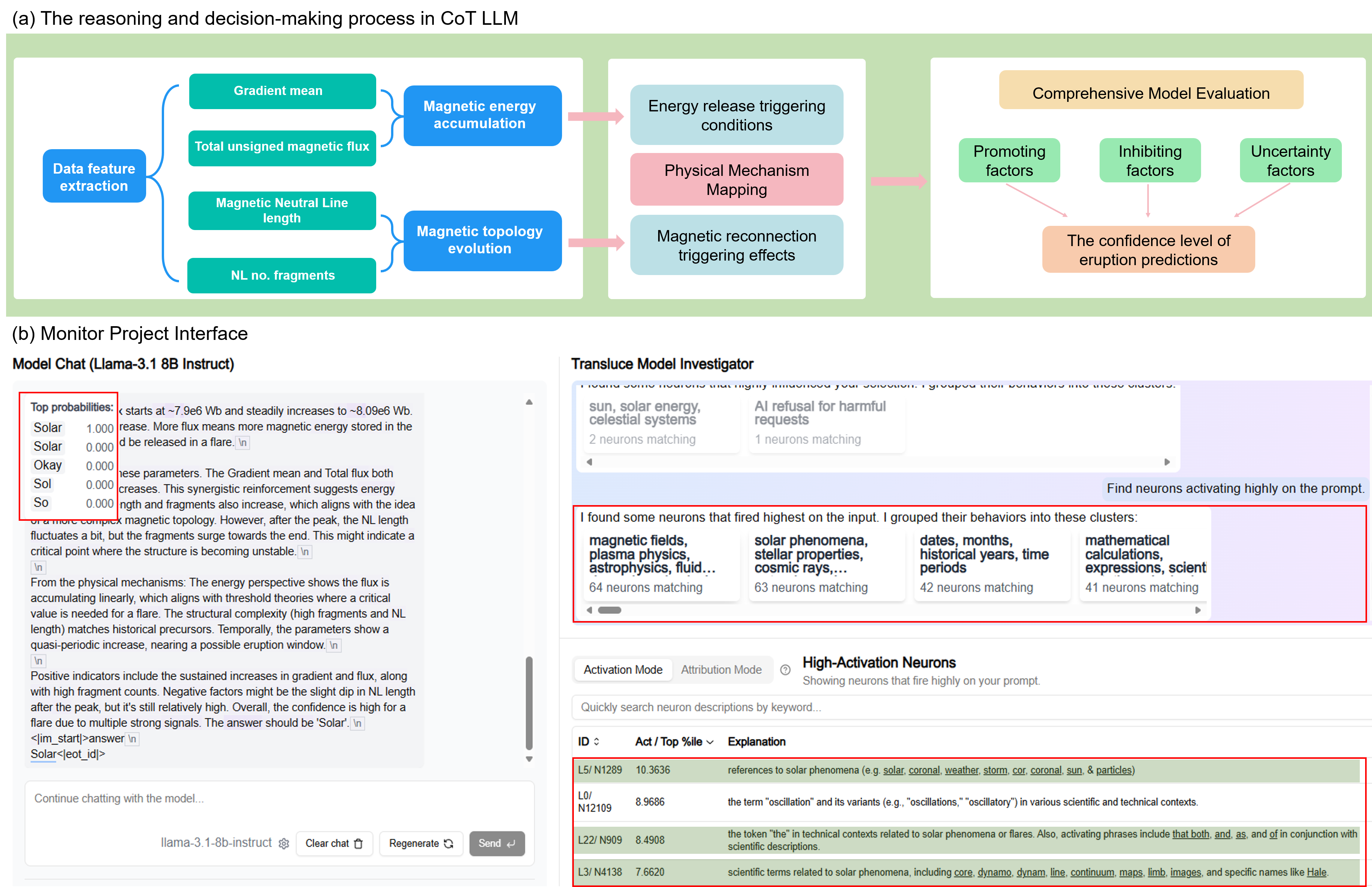}
  \caption{Explainability Experiment Analysis. (a) represents the reasoning process of the CoT LLM, while (b) illustrates the neuron features visualized through the Monitor project.}
  \label{fig:monitor_improvement}
\end{figure*}

As illustrated in Figure 4(a), the model's reasoning and decision-making process follows three stages: Initially, the model evaluates fluctuations in key parameters—such as Gradient mean and Total unsigned magnetic flux—to identify precursor patterns associated with the accumulation of magnetic energy, a critical indicator of potential solar flares. Simultaneously, it evaluates NL no. fragments and length fluctuations of magnetic neutral lines to characterize the evolution of magnetic topology. Subsequently, through synergistic parameter analysis, the model maps the extracted features to solar physical mechanisms, including energy release triggering conditions and magnetic reconnection triggering effects. This integration highlights the model’s ability to bridge data-driven insights with domain-specific knowledge. Finally, the model quantitatively assesses three categories of influencing factors—flare-promoting factors, inhibiting factors, and uncertainty factors—to derive the confidence level of eruption predictions. This process reveals that the LLM achieves solar flare prediction through a reasoning pathway involving data feature extraction, physical mechanism mapping, and comprehensive model evaluation, with interpretability mechanisms stemming from the implicit encoding of solar physics prior knowledge and explicit modeling of parameter associations.

Through a pre-compiled database of high-quality neuron descriptions, we tracked neural activity and found that solar-physics-related inputs significantly influenced specific neurons, which clustered into groups corresponding to categories such as solar phenomena and magnetic fields (As illustrated in the upper-right corner of Figure 4(b)). The lower-right panel of Figure 4(b) provides descriptions of individual highly active neurons, revealing that solar-related neurons were activated most frequently and dominated the activation patterns. These solar-associated neurons are prominently highlighted in the visualization. Another notable phenomenon is that not all displayed concepts are relevant to the predictions— for instance, neurons associated with "oscillation" were triggered; however, they exerted no significant influence on the final experimental outcomes. The prediction probability remained at 1, as highlighted in the red box on the left side of Figure 4(b), demonstrating the strong robustness of our algorithm. 

Explainability experiments employing knowledge distillation from the Deepseek-R1 model reveal the underlying flare prediction mechanisms within LLMs. The model identifies key magnetic precursors and maps them to physical processes such as energy accumulation and magnetic reconnection, enabling robust predictive outcomes through integrated assessment. Neuron activation tracking further demonstrates that the decision-making process preferentially engages solar physics-relevant neural. This explicit integration of domain-specific physical knowledge fundamentally distinguishes LLMs from conventional deep learning paradigms \cite{munikoti2024generalistmultimodalaireview}.

\subsection{Scaling Law for JW-Flare}

There is a notable Scaling Law phenomenon in the LLM field, which increases the amount of data and model parameters consistently improves performance. However, research on the application of LLMs in scientific domains is still limited, we conducted the Scaling Law experiment using 1B, 2B, 4B, and 7B models. Table 7 showed a progressive increase in the TSS metric, from 0.785 to 0.825, then to 0.935, and ultimately reaching 0.950. The results confirm that the Scaling Law phenomenon also exists in scientific domains. It can be inferred that LLMs with more parameters possess a greater number of relevant neurons during pre-training, and that the fine-tuning process preferentially activates domain-specific neurons, thereby enhancing predictive performance. 

\begin{table*}[h]
  \centering
  \renewcommand{\arraystretch}{2}
  \setlength\tabcolsep{2pt}
  \label{tab:Scaling_Law}
  \small
  \begin{tabular}{ccccccccccccc}
      \toprule
      Model & Level & TP & TN & FP & FN & TPR & TNR & HSS & ACC & Precision & F1 & \textbf{TSS} \\
      \midrule
      internVL2-1B   & $\geq X$ & 64 & 18400 & 470 & 15  & 0.810 & 0.975 & 0.203 & 0.974 & 0.120 & 0.209 & \textbf{0.785} \\
      Qwen2-vl-2B-instruct   & $\geq X$ & 69  & 17956  & 914  &  10  & 0.873  &  0.952 & 0.123  &  0.951 & 0.070 & 0.130  & \textbf{0.825}  \\
      internVL2-4B & $\geq X$ & 76  & 18368  & 502  &  3  & 0.962  &  0.973 & 0.226  &  0.973 & 0.132  & 0.231  &  \textbf{0.935} \\
      Qwen2-vl-7B-instruct & $\geq X$ & 79  & 17933 & 937  & 0  & 1.000 & 0.950 & 0.138 & 0.950 & 0.078 & 0.144 & \textbf{0.950}\\
      \bottomrule
  \end{tabular}
  \caption{Scaling Law for JW-Flare with a focus on forecasting X-class and larger solar flares. The bolded metrics represent key performance indicators, where higher values indicate stronger model performance. Due to the absence of 1B and 4B models in the Qwen2 series, InternVL2 was chosen for its performance, which is commonly considered on par with the Qwen2 models.}
\end{table*}

\subsection{Ablation Study}

To validate the effectiveness of the various components in JW-Flare, we conducted a series of ablation experiments, including JW-Flare (Data\_Balance), JW-Flare (Physics) and JW-Flare (Prompt). These experiments were designed to systematically assess the impact and rationale of sample balancing techniques, magnetic parameter input and prompt design of JW-Flare. This study enables a thorough evaluation of the impact of each module on the model's performance.

\begin{table*}[!htb]
  \centering
  \renewcommand{\arraystretch}{2}
  \setlength\tabcolsep{2pt}
  \label{tab:jwflare}
  \small
  \begin{tabular}{ccccccccccccc}
      \toprule
      Model & Level & TP & TN & FP & FN & TPR & TNR & HSS & ACC & Precision & F1 & \textbf{TSS} \\
      \midrule
      JW-Flare (Data\_Balance)   & $\geq X$ & 6 & 18797 & 73 & 71  & 0.078 & \textbf{0.996} & 0.073 & \textbf{0.992} & 0.076 & 0.077 & 0.074 \\
      JW-Flare (Physics)   & $\geq X$ & 75 & 17983 & 887 & 4  & 0.950 & 0.953 & 0.137 & 0.953 & \textbf{0.078} & \textbf{0.144} & 0.902 \\
      JW-Flare (Prompt) & $\geq X$ & 76 & 17633 & 1237 & 3  & 0.962 & 0.934 & 0.102 & 0.935 & 0.058 & 0.109 & 0.897 \\
      JW-Flare & $\geq X$ & 79  & 17933 & 937  & 0  & \textbf{1.000} & 0.950 & \textbf{0.138} & 0.950 & \textbf{0.078} & \textbf{0.144} & \textbf{0.950}\\
      \bottomrule
  \end{tabular}
  \caption{Ablation Study for JW-Flare with a focus on forecasting X-class and larger solar flares.. JW-Flare (Data\_Balance) model excluded data resampling and undersampling techniques; JW-Flare (Physics) model removed magnetic physical parameters as input; JW-Flare (Prompt) model replaced the step-by-step reasoning guidance and A/B options constraints with plain text output. The highest performance metrics are highlighted in bold.}
\end{table*}

As shown in Table 8, The JW-Flare (Data\_Balance) ablation experiment  confirms the necessity of addressing sample imbalance in flare prediction tasks, as evidenced by a drop in the TSS score to 0.074. The application of data resampling and undersampling techniques effectively mitigates the impact of data imbalance, ensuring that the model can more comprehensively learn the features of different categories. Moreover, the incorporation of magnetic parameters and efficient prompt design significantly enhance JW-Flare's predictive performance, fully leveraging the advantages of multimodal large models. Magnetic physical parameters provide more domain-specific knowledge, helping it better understand the physical context of solar activity, while efficient prompts guide the model's reasoning process, improving its performance on complex tasks. Experimental results show that JW-Flare outperforms all other models across key evaluation metrics, especially in TSS. These findings highlight the effectiveness of the proposed modules and confirm their individual contributions to overall performance improvement.

\section{Discussion}
This work presents JW-Flare -- the first accurate solar flare forecasting method based on open-source MLLMs. The model achieves perfect TPR (100\%) for X-class flares prediction, significantly outperforming traditional methods. The exceptional performance of JW-Flare stems from two  key factors: 1) The inherent next-token prediction capability of LLMs, combined with their pretraining-acquired knowledge of solar physics, establishes the foundation for addressing the complex challenge of flare forecasting. 2) Building upon this foundation, JW-Flare achieves high-precision prediction through the construction of a multimodal SFT dataset integrating parameterized physical texts with LoS magnetograms, and the implementation of domain-adapted prompt fine-tuning techniques. To validate the experimental design efficacy and uncover latent domain knowledge within LLMs, we conducted multi-faceted experimental evaluations. Explainability experiments reveal that our MLLM-based flare prediction method effectively activates pretrained solar physics knowledge during decision-making. Scaling law analysis demonstrates that domain-adapted MLLMs still observe the empirical principle of performance scaling with parameter size. Further ablation studies confirm the synergistic combination of multimodal supervised fine-tuning dataset construction and domain-adapted prompt fine-tuning techniques drives breakthrough predictive performance.

Looking ahead, JW-Flare’s performance can be further enhanced through three key strategies. First, incorporating long-term (decadal-scale) solar activity data may help capture more comprehensive temporal evolution patterns and finer details. Second, conducting extensive model training and validation across multiple temporal scales (ranging from hours to days) could improve the model's adaptability to multi-time-window prediction tasks. Third, developing hierarchical fusion strategies to integrate observational data from different layers of the solar atmosphere (e.g., photosphere, chromosphere, and corona) has the potential to enhance the model’s overall integrity and systematic analytical capabilities. These three approaches collectively focus on maximizing data utilization to refine the model. Furthermore, we will closely follow advancements in SOTA open-source LLMs (e.g., DeepSeek-R1, Qwen) to ensure alignment with the latest innovations in artificial intelligence.

This study proposes an artificial intelligence paradigm that decouples algorithmic innovation from domain expertise. By encapsulating complex computations within LLMs, the framework enables researchers to focus on curating high-quality scientific datasets—the critical determinant of downstream model performance. JW-Flare implements this approach through modular isolation of the SFT dataset construction process from the core predictive MLLM, achieving plug-and-play integration with cutting-edge MLLM architectures. Its performance exhibits progressive evolution alongside advancements in underlying MLLM capabilities. As high-quality scientific data resources advance synergistically with open-source LLM technologies, this paradigm establishes a fundamental framework for next-generation scientific discovery platforms.

\bibliography{aaai2026}


\end{document}